\documentclass[usenatbib,fleqn]{mnras}

\pdfoutput=1

\usepackage{amsmath}
\usepackage{amssymb}
\usepackage{bm}
\usepackage{color}
\usepackage{graphicx}
\usepackage[all]{hypcap}
\usepackage{hyperref}
\usepackage{multicol}
\usepackage{multirow}
\usepackage{natbib}
\usepackage{flushend}
\usepackage{txfonts}

\usepackage[capitalize,nameinlink]{cleveref}
\creflabelformat{equation}{#2#1#3}
\crefrangelabelformat{equation}{#3#1#4 to #5#2#6}

\hypersetup{colorlinks,
  linkcolor=red,
  filecolor=red,
  urlcolor=magenta,
  citecolor=blue}

\usepackage{array}
\newcolumntype{L}[1]{>{\raggedright\let\newline\\\arraybackslash\hspace{0pt}}m{#1}}
\newcolumntype{C}[1]{>{\centering\let\newline\\\arraybackslash\hspace{0pt}}m{#1}}
\newcolumntype{R}[1]{>{\raggedleft\let\newline\\\arraybackslash\hspace{0pt}}m{#1}}

\def\esd{\Delta\Sigma}
\def\lcdm{$\Lambda$CDM}
\def\kms{\mathrm{km\,s^{-1}}}

\def\mstar{m_\star}
\def\Msun{\mathrm{M}_\odot}

\def\percent{\%}

\def\sigmac{\Sigma_\mathrm{c}}

\def\galfit{\textsc{galfit}}

\def\meneacs{MENeaCS}
\def\planck{\textit{Planck}}

\def\bound{\mathrm{bound}}
\def\Mcl{M_\mathrm{cl}}
\def\percent{ per cent}
\def\reff{r_\mathrm{eff}}

\title[Weak lensing masses of UDGs]
      {A first constraint on the average mass of ultra diffuse galaxies from weak gravitational lensing}

\author[C.\ Sif\'on, R.\ F.\ J.\ van der Burg, H.\ Hoekstra, A.\ Muzzin \& R.\ Herbonnet]
  {
      Crist\'obal~Sif\'on$^{1,2}$,
      Remco~F.~J.~van~der~Burg$^3$,
      Henk~Hoekstra$^2$,
      Adam~Muzzin$^4$
      and
      \newauthor
      Ricardo~Herbonnet$^2$
\\
      $^1$Department of Astrophysical Sciences, Peyton Hall, Princeton University, Princeton, NJ 08544, USA\\
      $^2$Leiden Observatory, Leiden University, PO Box 9513, NL-2300 RA Leiden, Netherlands\\
      $^3$Laboratoire AIM, IRFU/Service d'Astrophysique - CEA/DSM - CNRS - Universit\'e Paris Diderot, B\^at. 709, CEA-Saclay, 91191 Gif-sur-Yvette Cedex, France\\
      $^4$Department of Physics and Astronomy, York University, 4700 Keele St., Toronto, Ontario, Canada, MJ3 1P3
  }

\pubyear{2016}

\begin{document}

\label{firstpage}
\pagerange{\pageref{firstpage}--\pageref{lastpage}}

\maketitle

\begin{abstract}
The recent discovery of thousands of ultra diffuse galaxies (UDGs) in nearby galaxy clusters has opened a new window into the process of galaxy formation and evolution. Several scenarios have been proposed to explain the formation history of UDGs, and their ability to survive in the harsh cluster environments. A key requirement to distinguish between these scenarios is a measurement of their halo masses which, due to their low surface brightnesses, has proven difficult if one relies on stellar tracers of the potential. We exploit weak gravitational lensing, a technique that does not depend on these baryonic tracers, to measure the average subhalo mass of 784 UDGs selected in 18 clusters at $z\leq0.09$. Our sample of UDGs has a median stellar mass $\langle\mstar\rangle=2\times10^8\,\Msun$ and a median effective radius $\langle\reff\rangle=2.8$ kpc. We constrain the average mass of subhaloes within 30 kpc to $\log m_{\rm UDG}(r<30\,\mathrm{kpc})/\Msun\leq10.99$ at 95\percent\ credibility, implying an effective virial mass $\log m_{200}/\Msun\leq11.80$, and a lower limit on the stellar mass fraction within 10 kpc of 1.0\percent. Such mass is consistent with a simple extrapolation of the subhalo-to-stellar mass relation of typical satellite galaxies in massive clusters. However, our analysis is not sensitive to scatter about this mean mass; the possibility remains that extreme UDGs reside in haloes as massive as the Milky Way.
\end{abstract}

\begin{keywords}
Gravitational lensing: weak -- Galaxies: evolution, formation, dwarf -- Galaxies: clusters: general -- Cosmology: dark matter
\end{keywords}

\section{Introduction}
\label{s:intro}

Large, low surface brightness galaxies have been known to exist both in the field \citep{dalcanton97} and in galaxy clusters \citep{impey88,turner93} for some time. A subset of these low surface brightness galaxies---dubbed ultra-diffuse galaxies (UDGs) by \cite{vandokkum15_coma}---has recently started receiving particular attention. UDGs stand out as relatively low-mass galaxies with large sizes, defined as having effective (or half-light) radii $\reff\geq1.5\,\mathrm{kpc}$, which are typical of $L^*$ galaxies with virial halo masses similar to that of the Milky Way, $M\sim10^{12}\,\Msun$. In contrast, their stellar masses are only of order $10^8\,\Msun$, two orders of magnitude lower than that of the Milky Way. A striking feature of UDGs is the fact that, despite their diffuse appearence and low stellar masses, they abound in and around massive galaxy clusters \citep{vandokkum15_coma,koda15,mihos15,yagi16,vdburg16,roman17_a168}. A number of UDGs have indeed been spectroscopically confirmed to be associated with the Coma cluster \citep{vandokkum15_spec,kadowaki17}.

Given their large sizes but low stellar masses, several hypotheses have been put forth to try to explain the unexpected survival of UDGs in massive clusters. In the initial discovery of UDGs in the Coma Cluster, \cite{vandokkum15_coma} suggested that UDGs may be failed galaxies, which fell into the cluster at early times after having used only a small fraction of their cold gas to form stars. Once part of a cluster, their remaining cold gas was removed and they were left as very dark matter dominated galaxies. \cite{yozin15} have shown that such a mechanism can in fact produce UDG-like galaxies in hydrodynamical simulations. Alternatively, \cite{amorisco16} and \cite{dicintio17} have suggested that the intrinsic properties of some dwarf galaxies are responsible for the formation of UDGs; if so, they should be abundant in the field as well as in clusters. In fact, a number of UDGs have been found in lower-density environments. The colours of these more isolated UDGs are consistent with those of field galaxies---they are blue, with ongoing star formation and a significant fraction of H \textsc{i} gas \citep[e.g.,][]{martinezdelgado16,merritt16,bellazzini17,roman17_groups,trujillo17}. There is strong evidence that the number of UDGs scales with the mass of the parent system from low-mass groups to massive clusters \citep{vdburg16,janssens17,roman17_groups}, but the interpretation of this observation is not straightforward \citep{roman17_groups}.

Among the properties of UDGs that may shed light on the models above and others, their total (i.e., halo) mass is a crucial one. Knowing the masses of UDGs would allow us to unambiguously rule out some classes of hypotheses, such as the `failed galaxy' hypothesis if their masses are small enough. In fact, there have been some attempts at estimating the masses of individual UDGs. \cite{beasley16_virgo} measured the velocity dispersion of globular clusters associated with VCC1287, a UDG in the Virgo Cluster, of $33_{-10}^{+16}\,\kms$ within 8.1 kpc, which suggests a virial mass\footnote{Throughout this work we use the term `virial mass' interchangeably with $m_{200}$. Here, $m_{200}$ is the mass within a radius $r_{200}$, within which the density is 200 times the mean density of the Universe at a given redshift. On a separate note, subhaloes cannot be physically assigned a virial mass, since they are embedded in the potential of the host cluster. However, referring to virial masses offers a convenient point for comparison. We therefore adhere to the use of virial masses in this discussion, but adopt a different mass definition in our analysis (see \Cref{s:results}). See also the discussion in \citet{sifon17_meneacs}.} $m_{200}\sim10^{11}\,\Msun$. Similar masses have been estimated for a few UDGs indirectly from the number of globular clusters \citep{beasley16_acs,peng16,vandokkum17}. All these measurements suggest that typical UDGs are dark matter dominated---perhaps `failed' galaxies---but have halo masses lower than those of $L^*$ galaxies. On the other hand, \cite{vandokkum16} estimated a virial mass $m\sim10^{12}\Msun$ from the stellar velocity dispersion and globular cluster count of DF44, a particularly large UDG  with $\reff=4.5\,\mathrm{kpc}$. Such mass is comparable to that of the Milky Way, suggesting that at the high-mass end UDGs may be as massive as $L^*$ galaxies.

Since all of the above mass estimations refer to single UDGs, it is not clear how they can be interpreted in the context of the UDG population, and therefore they are of limited use in distinguishing UDG formation hypotheses. Furthermore, all of these estimates of virial masses stem from order-of-magnitude extrapolations from measurements performed in all cases within 10 kpc. Instead, measurements of the \emph{total} masses of a representative population of UDGs are required to draw conclusions about their origin.

Weak gravitational lensing---the distortion of light from background galaxies by the mass distribution of objects closer to us---can provide direct measurements of the \emph{average} masses of UDGs. Gravitational lensing does not rely on faint, low-surface brightness baryonic tracers of the mass in galaxies, and therefore provides a clear advantage over stellar dispersion measurements or globular cluster counts when trying to estimate UDG masses. For instance, the globular clusters identified by \cite{beasley16_acs} have magnitudes $m_\mathrm{F814W}\gtrsim26$ in the AB system, for a UDG \emph{in the Coma Cluster}. Evidently, such an analysis would be observationally expensive for more distant objects.

The feasibility of weak lensing for measuring the total masses of satellite galaxies has already been demonstrated on a variety of datasets \citep[e.g.,][]{natarajan02,natarajan09,gillis13,li14,li16,sifon15_kids,sifon17_meneacs,niemiec17}. Here, we present the first direct constraints on the average masses of UDGs using measurements of weak lensing produced by UDGs in a sample of 18 clusters at $z\leq0.09$ taken from the Multi-Epoch Nearby Cluster Survey \citep[\meneacs,][]{sand12}. Throughout this work we adopt a flat \lcdm\ model with $\Omega_\mathrm{m}=0.315$ and $H_0=70\,\mathrm{km\,s^{-1}Mpc^{-1}}$, consistent with the latest measurements of the cosmic microwave background by the \planck\ satellite \citep{planck15xiii}.

\section{Data analysis}
\label{s:data}

\subsection{UDG sample}
\label{s:udg}

\meneacs\ is a multi-epoch survey of 57 clusters at $z\leq0.15$ carried out with Megacam at the Canada-France-Hawaii Telescope \citep{sand12}. The data reduction is described in detail by \cite{vdburg13,vdburg15}. The resulting full width at half maximum of the point spread function is less than $1''$ for all clusters in the sample, and the photometric zero points have been calibrated to about 0.01 mag. We consider 18 of those 57 clusters selected to be affected by low Galactic extinction $A_r\leq0.2$ (Herbonnet et al.\ in prep.), and located at $z\leq0.09$ where we can still identify most UDGs given their surface brightness. These clusters have halo masses $M_{200}\gtrsim10^{14}\,\Msun$, as determined from both galaxy velocity dispersions \citep{sifon15_cccp} and weak lensing measurements (Herbonnet et al.\ in prep.). We list the cluster sample in \Cref{t:clusters}.

Following the original definition by \cite{vandokkum15_coma}, \cite{vdburg16} identified UDGs in a subset of our clusters as galaxies with average surface brightnesses \citep[as opposed to central surface brightness used e.g., by][]{vandokkum15_coma} within one effective radius of $24.0\leq \langle\mu(r,\reff)\rangle\leq26.5$ and effective radii $1.5\leq\reff/\mathrm{kpc}\leq7.0$; these parameters were measured from \galfit\ \citep{peng02,peng10} single-S\'ersic fits to the light distribution of each galaxy. In order to have a sample that is as pure as possible but still be able to obtain a large enough number of UDG candidates, \cite{vdburg16} considered eight clusters at $0.04\leq z \leq0.07$ and at galactic latitude $\lvert b \rvert \geq 25^\circ$. The study of \cite{vdburg16} is unique in that they did not select UDGs by visual inspection after the automatic selection based on structural parameters. While this has the disadvantage that the sample may be (and in fact is, as we discuss below) contaminated by artefacts of various kinds, it allows for an objective, statistically sound study of their properties, after accounting for the expected number of such objects (both real and artefacts) in control fields. This contamination however can significantly alter lensing measurements in a way that may not be fully captured by subtracting the signal from a control sample, and we therefore refine (and expand) the sample of \cite{vdburg16} for weak lensing measurements. We extend the cluster sample to $z\leq0.09$ in order to increase the number of UDGs in our sample. We also impose stricter size cuts on the UDG sample: $\reff\geq2.0\,\mathrm{kpc}$ for $z\leq0.07$ and $\reff\geq3.0\,\mathrm{kpc}$ for higher redshifts, as we find that these cuts significantly reduce the contamination of the sample. Since most of the lensing signal is expected to be produced by the more luminous UDGs in the first place (which are on average larger), these two extra steps ensure a cleaner signal without significantly losing information. As \cite{vdburg16}, we include all UDG candidates within each cluster's $r_{200}$ \citep{sifon15_cccp}. We find that restricting our sample to smaller radii to potentially reduce contamination has no impact on our results.

At this point our sample of UDG candidates contains two kinds of contaminants, which we refer to as artefacts and interlopers. The former are instrumental or astrophysical contaminants such as stellar spikes, blended objects or resolved features within spiral arms, while the latter are true galaxies but are not part of the clusters, and therefore complicate the interpretation of our measurements. We discard artefacts by visual inspection, but we cannot remove interlopers without redshift information and they therefore enter our lensing analysis. We discuss the possible effects of interlopers in \Cref{s:masses}.

As mentioned above, we visually inspect all UDG candidates and only keep high-confidence ones. We also reject blended galaxies through visual inspection. These galaxies are not artefacts as such, nor are they necessarily interlopers (i.e., they may well be part of the cluster), but their best-fit \galfit\ parameters are biased since they are assumed to be a single object, and therefore complicate the interpretation of our results. Therefore, we classify a UDG candidate as high-confidence if its best-fit \galfit\ model can be regarded free of contamination from neighbouring galaxies (i.e., if a model-subtracted image shows no significant residuals). We do not make any further visual selection based on surface brightness, colour or morphology. This is an important distinction from the strategy of statistical studies of UDG numbers such as the one by \cite{vdburg16}. These studies should avoid subjective selections \citep[and can instead properly account for contamination with the aid of control samples, see][]{vdburg16}, but this is an important step in our analysis that allows for a clean interpretation of the lensing signal. Since our sample contains both artefacts and interlopers, subtraction of a control signal is not guaranteed to allow a robust interpretation. Based on our visual inspection, we reject 14\percent\ of the automatically-selected candidates. Some of these rejected objects are true artefacts, but most of them are blended galaxies. 

The final number of UDG candidates per cluster is listed in \Cref{t:clusters}, along with the expected number of interlopers. The latter have been calculated by multiplying the density of objects in the control fields that pass our automated and visual selections by the total area analyzed in each cluster \citep[see][]{vdburg16}. We find a total of 784 UDG candidates with an estimated average interloper fraction of 40\percent. We discuss the implications of this contamination for our analysis in \Cref{s:masses}.

\begin{figure}
  \centerline{\includegraphics[width=\linewidth]{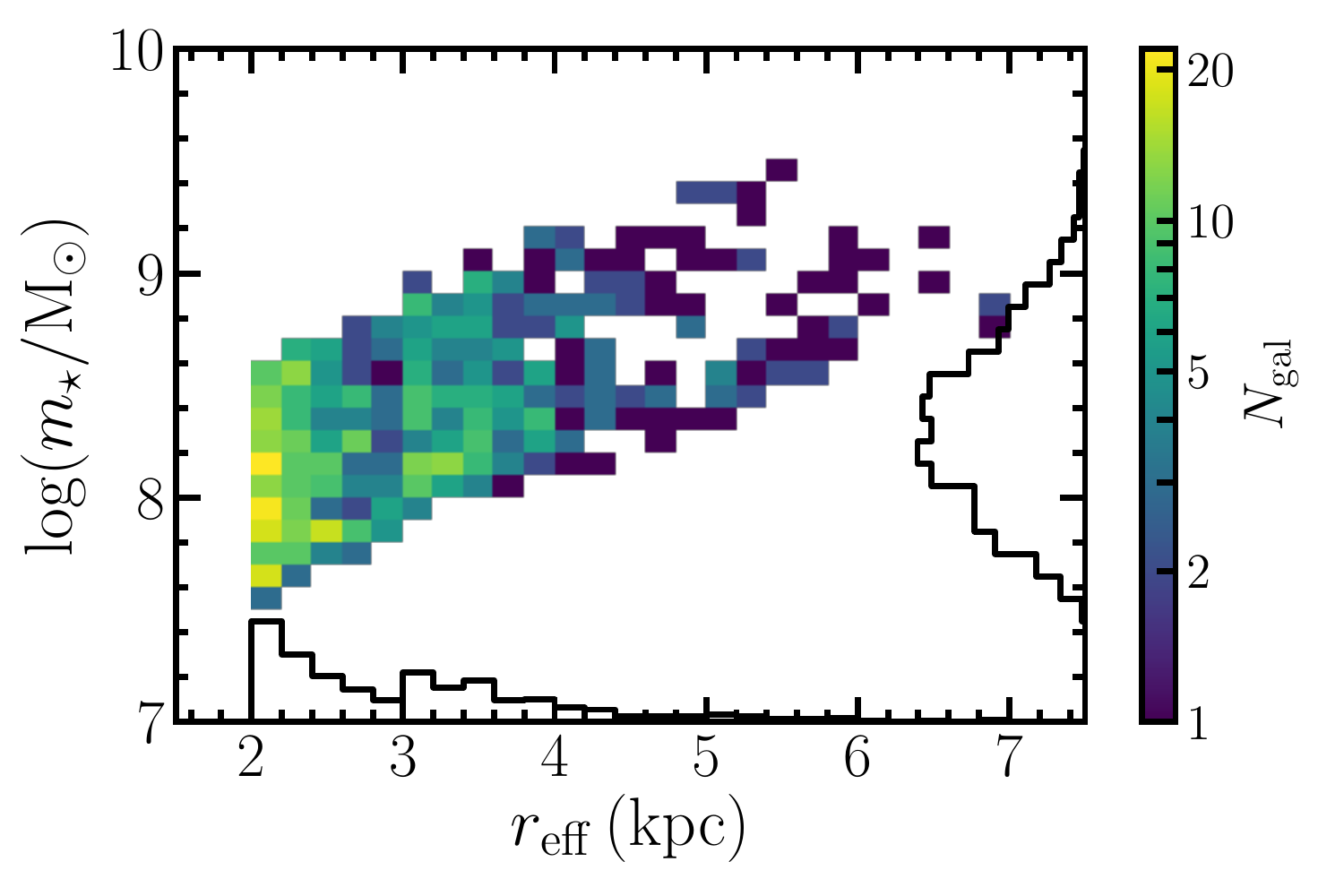}}
  \caption{One- and two-dimensional distributions of effective radius and stellar mass of all UDG candidates that pass our visual inspection. The colour scale shows the number of UDGs in a logarithmic scale. We only include UDG candidates with $\reff\geq2$ kpc at $0.04<z\leq0.07$ and with $\reff\geq3$ kpc at $z>0.07$.}
  \label{f:hist}  
\end{figure}

We estimate stellar masses using the $g-r$ colours assuming no dust, solar metallicity, and the initial mass function of \cite{chabrier03}, assuming that all UDGs are part of the corresponding cluster. We show the distributions of stellar mass and effective radii of our final sample of UDG candidates in \Cref{f:hist}. They have a median stellar mass $\langle\mstar\rangle=2\times10^8\,\Msun$, and a median effective radius $\langle\reff\rangle=2.8\,\mathrm{kpc}$. The median S\'ersic index is $\langle n_\mathrm{Sersic} \rangle = 1.7$; this is larger than the typical index of UDGs in Coma \citep[$\langle n_\mathrm{Sersic}\rangle\sim0.8$, e.g.,][]{koda15,yagi16}, which may partly result from the selection in mean surface brightness within $\reff$ rather than central surface brighntess \citep[see the discussion in][]{vdburg16}.

\begin{table}
 \centering
\caption{Cluster sample. Clusters masses, $M_{200}$, refer to the dynamical masses estimated by \citet{sifon15_cccp}, and $r_{200}$ are the radii containing such masses. Note that in this table the overdensity is defined with respect to the critical density of the Universe. The last column lists the number of good UDG candidates after cleaning the sample by visual inspection, without correcting for further contamination, and the estimated number of interlopers in parentheses.}
\label{t:clusters}
\begin{tabular}{l c r@{ $\pm$ }l c r@{ }r}
\hline\hline
 \multirow{2}{*}{Cluster} & \multirow{2}{*}{Redshift} & \multicolumn{2}{c}{$M_{200}$} & $r_{200}$ & \multicolumn{2}{c}{Number} \\
         &          & \multicolumn{2}{c}{$(10^{14}\Msun)$} & $(\mathrm{Mpc})$ & \multicolumn{2}{c}{of UDGs}  \\[0.5ex]
\hline
A85        & 0.055 & 10.2 & 1.8 & $2.0\pm0.1$ &  97 & (33) \\
A119       & 0.044 &  7.4 & 1.2 & $1.8\pm0.1$ &  82 & (17) \\
A133       & 0.056 &  5.5 & 1.6 & $1.7\pm0.2$ &  78 & (27) \\
A780       & 0.055 &  7.2 & 2.7 & $1.8\pm0.2$ &  57 & (25) \\
A1650      & 0.084 &  4.5 & 0.9 & $1.5\pm0.1$ &  16 &  (9) \\
A1651      & 0.085 &  8.0 & 1.3 & $1.9\pm0.1$ &  35 & (13) \\
A1781      & 0.062 &  0.6 & 0.3 & $0.8\pm0.1$ &  20 &  (9) \\
A1795      & 0.063 &  5.0 & 0.9 & $1.6\pm0.1$ & 109 & (30) \\
A1991      & 0.059 &  1.9 & 0.5 & $1.2\pm0.1$ &  35 & (14) \\
A2029      & 0.078 & 16.1 & 2.5 & $2.4\pm0.1$ &  34 & (19) \\
A2033      & 0.080 &  7.7 & 2.0 & $1.8\pm0.2$ &  25 & (13) \\
A2065      & 0.072 & 14.4 & 2.5 & $2.3\pm0.1$ &  19 & (13) \\
A2142      & 0.090 & 13.8 & 1.2 & $2.2\pm0.1$ &  62 & (20) \\
A2495      & 0.079 &  4.1 & 0.8 & $1.5\pm0.1$ &  13 &  (6) \\
A2597      & 0.083 &  3.5 & 2.0 & $1.4\pm0.3$ &  20 &  (8) \\
A2670      & 0.076 &  8.5 & 1.2 & $1.9\pm0.1$ &  26 & (11) \\
MKW3S      & 0.045 &  2.6 & 0.5 & $1.3\pm0.1$ &  23 &  (8) \\
ZWCL1215   & 0.077 &  7.7 & 1.7 & $1.9\pm0.1$ &  33 & (11) \\
[0.5ex]
\hline
\end{tabular}
\end{table}

\subsection{Weak lensing measurements}

Our weak lensing analysis is identical in methodology to that presented in \cite{sifon17_meneacs}, which follows closely the cluster lensing analyses of \cite{hoekstra15} and Herbonnet et al.\ (in prep.). The weak lensing signal is measured as an average tangential alignment, or shear, $\gamma_\mathrm{t}$, of ``source'' galaxies in the background of the lenses (in this case, UDGs) using the moments-based KSB algorithm \citep{kaiser95,luppino97,hoekstra98}. The shear is related to the excess surface density, $\esd$, 
\begin{equation}\label{eq:esd}
 \esd \equiv \bar\Sigma(<R) - \bar\Sigma(R) = \sigmac\gamma_\mathrm{t},
\end{equation}
where $\bar\Sigma(<R)$ and $\bar\Sigma(R)$ are the average surface densities within a projected radius\footnote{As a convention, we list two-dimensional distances with upper case $R$ and three-dimensional distances with lower case $r$.} $R$ and within a thin annulus around $R$, respectively, and $\sigmac$ is a geometric factor accounting for the lensing efficiency,
\begin{equation}\label{eq:sigmac}
 \sigmac \equiv \frac{c^2}{4\pi G} \frac{D_\mathrm{s}}{D_\mathrm{l}D_\mathrm{ls}}\,,
\end{equation}
where $D_\mathrm{s}$, $D_\mathrm{l}$ and $D_\mathrm{ls}$ are the angular diameter distances to the source, to the lens, and between the lens and source, respectively. As described in Herbonnet et al.\ (in prep.), we calculate $D_\mathrm{s}$ and $D_\mathrm{ls}$ by matching the magnitude distribution of the source sample to the 30-band COSMOS2015 photometric redshifts \citep{laigle16}, appropriately adjusted to the noise level of each \meneacs\ cluster.

We construct the source sample in an identical way to Herbonnet et al.\ (in prep.) and \cite{sifon17_meneacs}. We do not apply any colour cuts to the sample, but correct the lensing measurements for contamination of our source sample by faint cluster members by comparing the source density in \meneacs\ clusters to that measured with our pipeline on archival CFHT Megacam data which do not contain any clusters in the field of view \citep[Herbonnet et al.\ in prep.; see also][]{hoekstra15}. The correction accounts for obscuration of background sources by bright cluster galaxies, the latter calculated using custom image simulations of both the cluster and source galaxies \citep{sifon17_meneacs}. Since UDGs are too faint to produce any significant obscuration \citep{sifon17_meneacs}, we simply apply an average correction for cluster member contamination (`boost correction') as a function of cluster-centric distance following the procedure outlined in Herbonnet et al.\ (in prep.). This correction results in approximately a 20\percent\ increase in the shear at 300 kpc from the cluster centre, and is on average $\approx10$\percent\ for our source sample, well within our statistical uncertainties. The uncertainty in this correction is at the level of 1\percent. In addition, \cite{sifon17_meneacs} used another set of image simulations to estimate the effect of cluster galaxies on the shapes measured for source galaxies. The uncertainty on this effect is at the level of 1\percent\ for the massive satellites analyzed by \cite{sifon17_meneacs}, and is negligible for UDGs. Finally, Herbonnet et al.\ (in prep.) estimated the uncertainty in \Cref{eq:sigmac} due to uncertainties in the mean source redshift to be below 2\percent\ for each cluster; the uncertainty on our stacked measurement is well below 1\percent.

\section{Weak lensing by Ultra Diffuse Galaxies}
\label{s:results}

\subsection{Model for the UDG lensing signal}
\label{s:model}

As discussed in detail by \cite{yang06} and \cite{sifon15_kids}, the satellite lensing signal can be described as the sum of terms accounting for the stellar, subhalo and host halo contributions,
\begin{equation}\label{eq:esd_tot}
  \Delta\Sigma = \Delta\Sigma_\star + \Delta\Sigma_\mathrm{sub} + \Delta\Sigma_\mathrm{host};
\end{equation}
the latter two dominate at different projected separations and can therefore be cleanly separated, while the stellar component dominates at scales typically smaller than are observationally accessible and does not contribute significantly to the measured signal. As in \cite{sifon15_kids,sifon17_meneacs}, we model the average density profile of both UDGs and their host clusters as NFW profiles \citep{nfw95},
\begin{equation}
  \rho_\mathrm{NFW} = \frac{\delta_c\rho_\mathrm{m}}{r/r_\mathrm{s}\left(1+r/r_\mathrm{s}\right)^2},
\end{equation}
where 
\begin{equation}
  \delta_c = \frac{200}{3} \frac{c^3}{\ln(1+c) - c/(1+c)}
\end{equation}
and $\rho_\mathrm{m}$ is the mean Universal density at the average redshift of our sample, $\langle z\rangle=0.062$. The NFW concentration, $c$, and scale radius, $r_\mathrm{s}$, are related through $c=r_{200}/r_\mathrm{s}$, where $r_{200}$ is the radius enclosing a density $\rho(<r_{200})=200\rho_\mathrm{m}$; the mass contained within $r_{200}$ is referred to as $m_{200}$. The stellar contribution to the lensing signal from the UDG as a point mass with an amplitude fixed to the average stellar mass of our sample, $\langle \mstar \rangle = 2.0\times10^8\,\Msun$, $\Delta\Sigma_\star = \langle m_\star \rangle/(\pi R^2)$.

The choice of a full NFW profile may not be entirely accurate for subhaloes (having suffered tidal stripping), but the lensing signals produced by density profiles with varying truncation strengths are indistinguishable with current errorbars \citep{sifon15_kids}. We therefore adopt a full NFW profile for simplicity, and because i) we focus our discussion around the mass within a fixed physical aperture which is relatively insensitive to the choice of density profile, and ii) it allows for a consistent comparison with other studies in \Cref{s:discussion}.

Following \cite{sifon17_meneacs}, we define the mass of subhaloes hosting UDGs, $m_\mathrm{bound}$, as the bound subhalo mass---that is, the mass within the region where the subhalo density is above the background density of the cluster. This definition is similar to that employed by some real-space subhalo finders such as {\sc subfind} \citep{springel01_cluster}, and is therefore a useful definition for comparisons with numerical and hydrodynamical simulations. In order to estimate the background density, we take the expected value for the three-dimensional cluster-centric distance,
\begin{equation}
 \langle r_\mathrm{UDG} \rangle = 
    \int_{0.15c}^c d\chi\,\chi\,\rho(\chi, c) \,
        \bigg/ \int_{0.15c}^c d\chi\,\rho(\chi, c)
     = 0.38r_\mathrm{200,cl}
    \,,
\end{equation}
where $\rho(\chi, c)$ is the NFW profile with the radial variable $\chi$ in units of the scale radius, $r_\mathrm{s} \equiv r_{200}/c$. Here, we fix a concentration $c=2$, which is a good fit to the radial distribution of typical satellite galaxies in \meneacs\ clusters \citep{vdburg15}, and also of UDGs at distances $r_\mathrm{sat}\gtrsim0.15\,r_\mathrm{200,cl}\sim300\,\mathrm{kpc}$ \citep{vdburg16}. We account for the radial distribution of UDGs in our calculation of the contribution from the host clusters, which smooths the transition between scales dominated by the subhalo and by the host halo \citep[see section 3.2 in][]{sifon15_kids}.

We find that the average concentrations of both clusters and UDGs are poorly constrained, but that they have little impact on our results. Due to the effect of tidal stripping, subhaloes are more centrally-concentrated than host haloes, and we expect that the same applies to UDGs. In fact, their shallow stellar profiles and their presence down to a few hundred kpc in the most massive clusters \citep[e.g.,][]{vdburg16} suggest that either their gravitational potentials are highly centrally concentrated or that UDGs have large total masses. We therefore fix the concentrations of clusters and UDGs through the mass-concentration relations of central haloes from \cite{dutton14} and of subhaloes from \cite{moline17}, respectively, both derived using N-body simulations.

We fit this model to the measurements using the affine-invariant Markov Chain Monte Carlo (MCMC) ensemble sampler \texttt{emcee} \citep{goodman10,foreman13}, fully accounting for the data covariance as described in \cite{sifon15_kids,sifon17_meneacs}. We adopt uniform priors for the masses (in logarithmic space) in the following ranges:
\begin{equation}
\begin{split}
  8.4 &\leq \log(m_\mathrm{200,UDG}/\Msun) \leq 13.0 \\
 14.0 &\leq \log(M_\mathrm{200,cl}/\Msun) \leq 16.0
 \,.
\end{split}
\end{equation}

The lower limit on the mass of UDGs is set by the average stellar mass of our sample. 
Note that the prior for the UDG masses refers to the total mass, which is then separated into a point mass and NFW components (i.e., a model with $\log(m_\mathrm{200,UDG}/\Msun)=8.3$ would have only a point mass contribution). Host cluster masses are well constrained (see \Cref{s:masses}) and therefore a more informative prior based on dynamical or weak lensing masses does not improve our constraints on UDG masses.

\subsection{The average mass of UDGs}
\label{s:masses}

\begin{figure}
 \centerline{\includegraphics[width=0.95\linewidth]{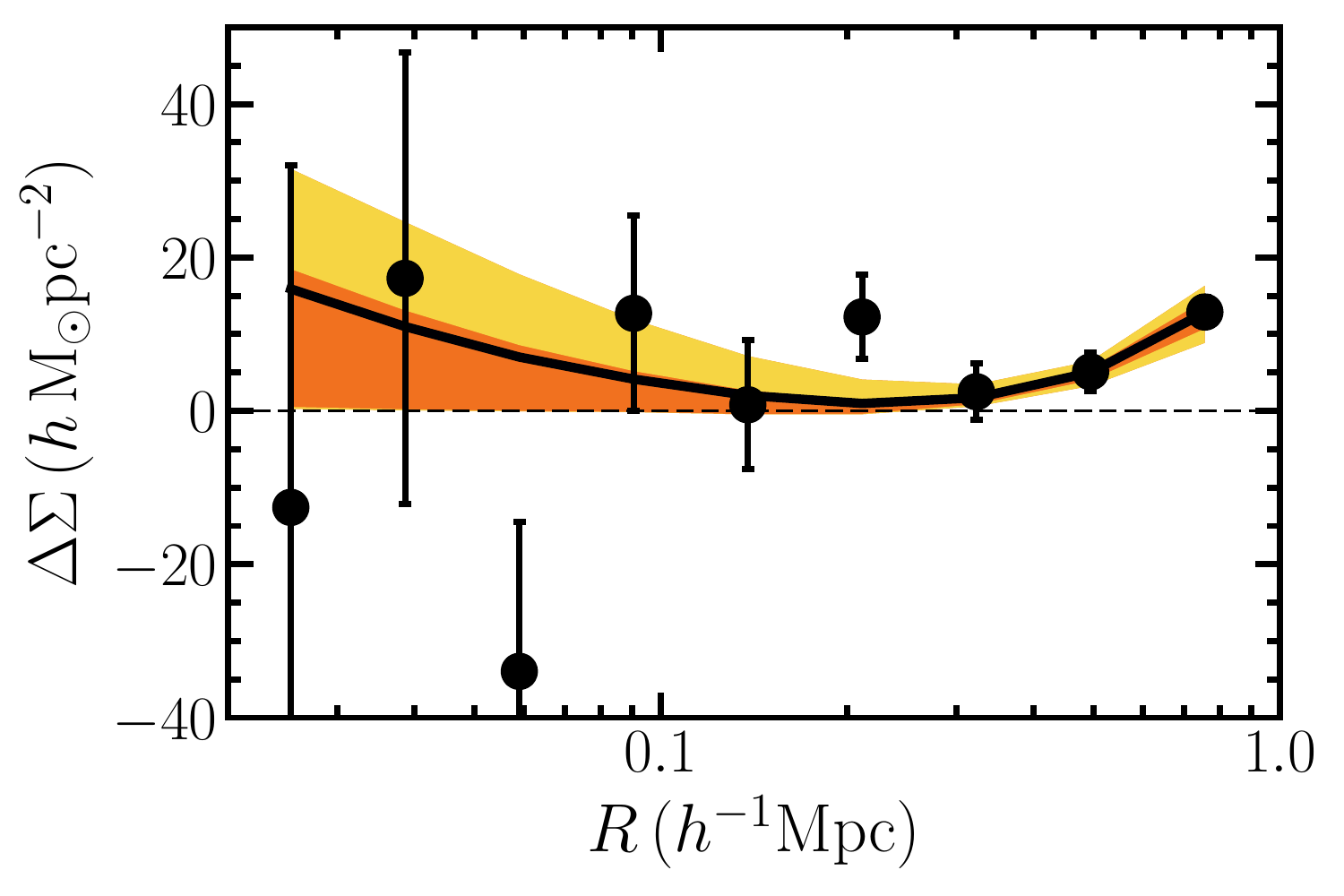}}
\caption{Weak lensing signal of ultra-diffuse galaxies in \meneacs\ clusters at $z\leq0.09$. Orange and yellow regions show the 68 and 95\percent\ credible intervals from the MCMC sampling, while the black line shows the best-fit model.}
\label{f:esd}
\end{figure}

\begin{figure}
 \centerline{\includegraphics[width=0.95\linewidth]{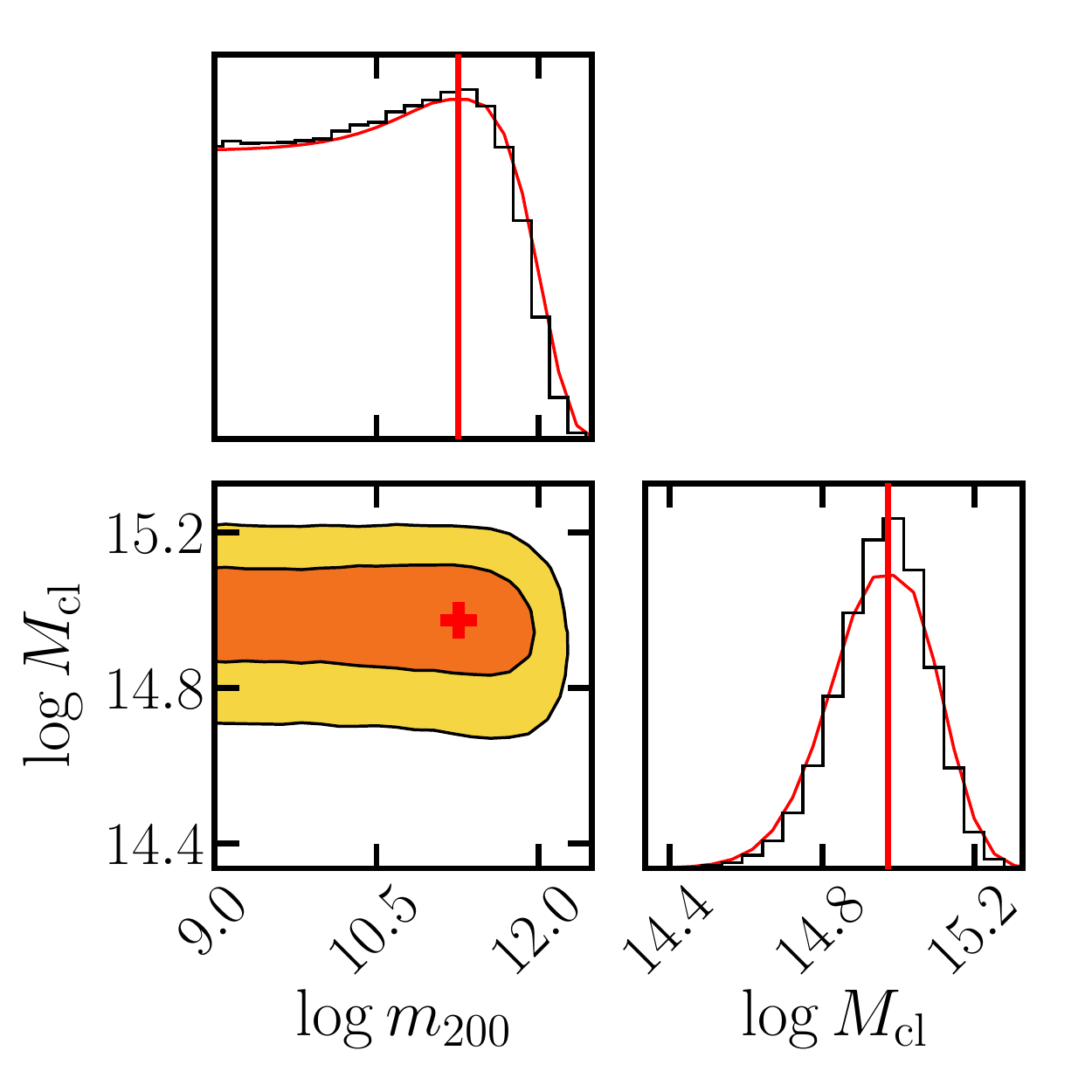}}
\caption{Marginalized and joint posterior distributions of the two free parameters in our model, namely the UDG and cluster average masses, both defined as the masses within their respective $r_{200}$ (see \Cref{s:masses} for a discussion of the definition of UDG masses). Bottom-left panel: contours show 68 and 95\percent\ credible regions and the red cross shows the maximum likelihood value. Top-left and bottom-right: black lines show the posterior distributions, while red lines show the projection of the maximum likelihood surface onto each parameter; the red lines show the maximum likelihood estimates.}
\label{f:corner}
\end{figure}

We show in \Cref{f:esd} the lensing signal of our sample of UDGs, along with the best-fit model. The UDG lensing signal is consistent with zero within $1\sigma$. The 95\percent\ credibility upper limit for the bound mass is $\log m_\bound/\Msun\leq11.69$ within a binding radius $r_\bound\leq106\,\mathrm{kpc}$. Note that $m_\bound$ depends on both the inferred subhalo $m_{200}$ and host cluster $M_{200}$ at every step in the MCMC chain, and the variation in these two parameters drives our constraint on $m_\bound$. The fact that the maximum likelihood estimate is so close to the 68\percent\ credible interval boundary, as shown in \Cref{f:esd}, is probably due to the fact that data points at $0.1<R/\mathrm{Mpc}<0.2$ carry the most weight in terms of the subhalo constribution to $\Delta\Sigma$, and the slope of the curve is then set by the use of a fixed mass-concentration relation. As shown in \Cref{f:corner}, the posterior distribution for the average UDG mass closely follows the likelihood distribution, and the peaks of both distributions coincide; the location of the best-fit line in \Cref{f:esd} is not an artefact of the sampling method. Ideally, we would measure the lensing signal of the control sample defined in \Cref{s:udg} to establish a `background' signal. However, our control sample contains too few objects to measure a meaningful signal, so this test cannot be carried out.

In the context of our model, we obtain a 95\percent\ credible upper limit of the `virial' subhalo masses of UDGs of $\log m_{200}/\Msun\leq11.80$ within a radius $r_\mathrm{200,UDG}\leq0.20\,\mathrm{Mpc}$. The best-fit average host cluster mass is $\log\Mcl/\Msun=14.96_{-0.12}^{+0.11}$ (68\percent\ credible interval), broadly consistent with determinations of the masses of these clusters from galaxy dynamics \citep{sifon15_cccp} and cluster weak lensing measurements (Herbonnet et al.\ in prep.).

For ease of comparison with recent measurements of UDG masses from stellar dynamics, we also quote the mass implied by our NFW model within fixed physical radii,
\begin{equation}
  \begin{split}
    \log m(r<10\,\mathrm{kpc})/\Msun &\leq 10.29 \, , \\
    \log m(r<20\,\mathrm{kpc})/\Msun &\leq 10.75 \, , \\
    \log m(r<30\,\mathrm{kpc})/\Msun &\leq 10.99 \, ,
  \end{split}
\end{equation}
all at 95\percent\ credibility. As can be deduced from the data points in \Cref{f:esd}, the first two values result from an extrapolation of our data, but are useful as a point of comparison. We adopt the mass within 30 kpc for most of the discussion that follows since we directly probe this radius with our measurements, and $m(<30\,\mathrm{kpc})$ is fairly insensitive to the choice of density profile given current uncertainty levels.

Weak gravitational lensing measures the total mass, irrespective of its nature. The median stellar mass of UDGs in our sample is $\langle m_\star \rangle = 2.0\times10^8\,\Msun$. On average, UDGs in clusters could have a stellar-to-total mass fraction as low as 1\percent\ within 10 kpc and 0.4\percent\ within $r_\bound$, although our results are also consistent with the masses of UDGs being dominated by their stellar content.
As discussed in \Cref{s:data}, while we have ensured that there are no true artefacts in our sample (including instrumental artefacts and blended galaxies), there is probably a non-zero fraction of interlopers---bonafide galaxies that pass all our cuts but are not part of the clusters. Our automatic selection of UDGs based on mean surface brightness finds galaxies that tend to me more centrally-concentrated than a selection based on central surface brightness. Coupled with the the low redshift of our clusters, this means that interlopers in our sample are more likely to be background, rather than foreground, galaxies. If this is the case, then they must be intrinsically brighter than we are assuming, so they presumably reside in more massive haloes, and bias our mass estimate \emph{high}. Therefore, the net effect of interlopers in our analysis is to favour a detection rather than to hide one. We therefore do not attempt to correct for the presence of interlopers in our analysis, and regard our upper limit on the average UDG mass a conservative one.

\subsection{UDG masses in context}

\begin{figure}
  \centerline{\includegraphics[width=0.95\linewidth]{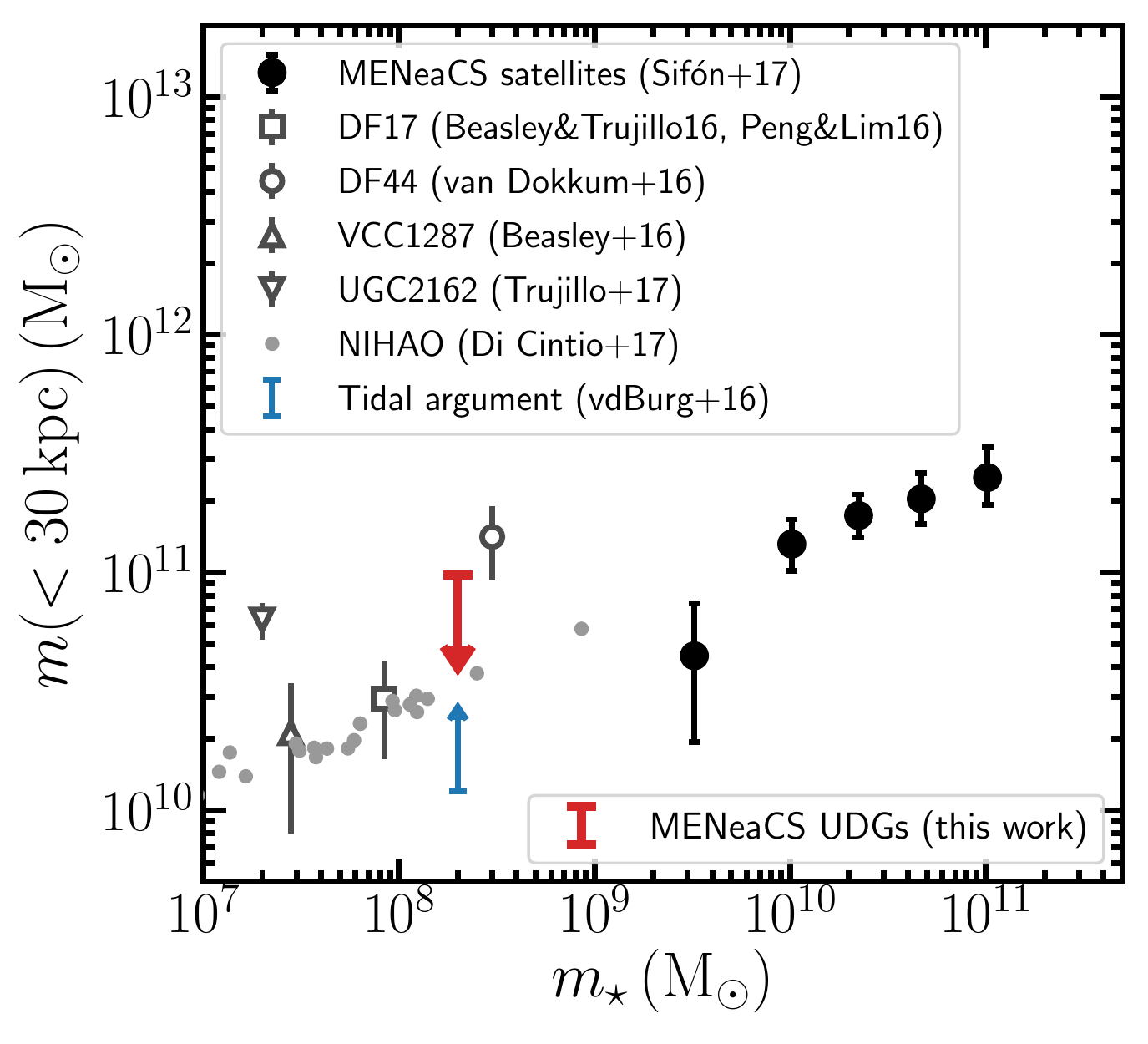}}\caption{Weak lensing--inferred 95\percent\ credible upper limit on the average halo mass of MENeaCS UDGs within 30 kpc. We also show for comparison the weak lensing measurements of the masses within 30 kpc of \meneacs\ satellite galaxies by \citet{sifon17_meneacs}, and  extrapolations to 30 kpc of mass measurements of individual UDGs from the literature including UDGs in the NIHAO simulations \citep{dicintio17} and a lower limit on the average mass based on tidal arguments by \citet{vdburg16}. Uncertainties for UDG masses from the literature have been scaled directly from those reported in each work and do not account for uncertainties associated with the extrapolation itself.}
\label{f:shmr30kpc}
\end{figure}

We compare our constraint on the average mass of UDGs to the total masses within 30 kpc of satellite galaxies with $\mstar>9\times10^9\,\Msun$ in \meneacs\ clusters by \cite{sifon17_meneacs} in \Cref{f:shmr30kpc}. Our constraints do not allow the average UDG to deviate significantly from a simple extrapolation of the total-to-stellar mass relation of satellites to lower stellar masses. However, our measurements cannot constrain the scatter in the halo mass of UDGs at fixed stellar mass, which may result in extreme objects such as DF44 \citep{vandokkum16} having significantly larger halo masses.

\begin{figure}
  \centerline{\includegraphics[width=0.95\linewidth]{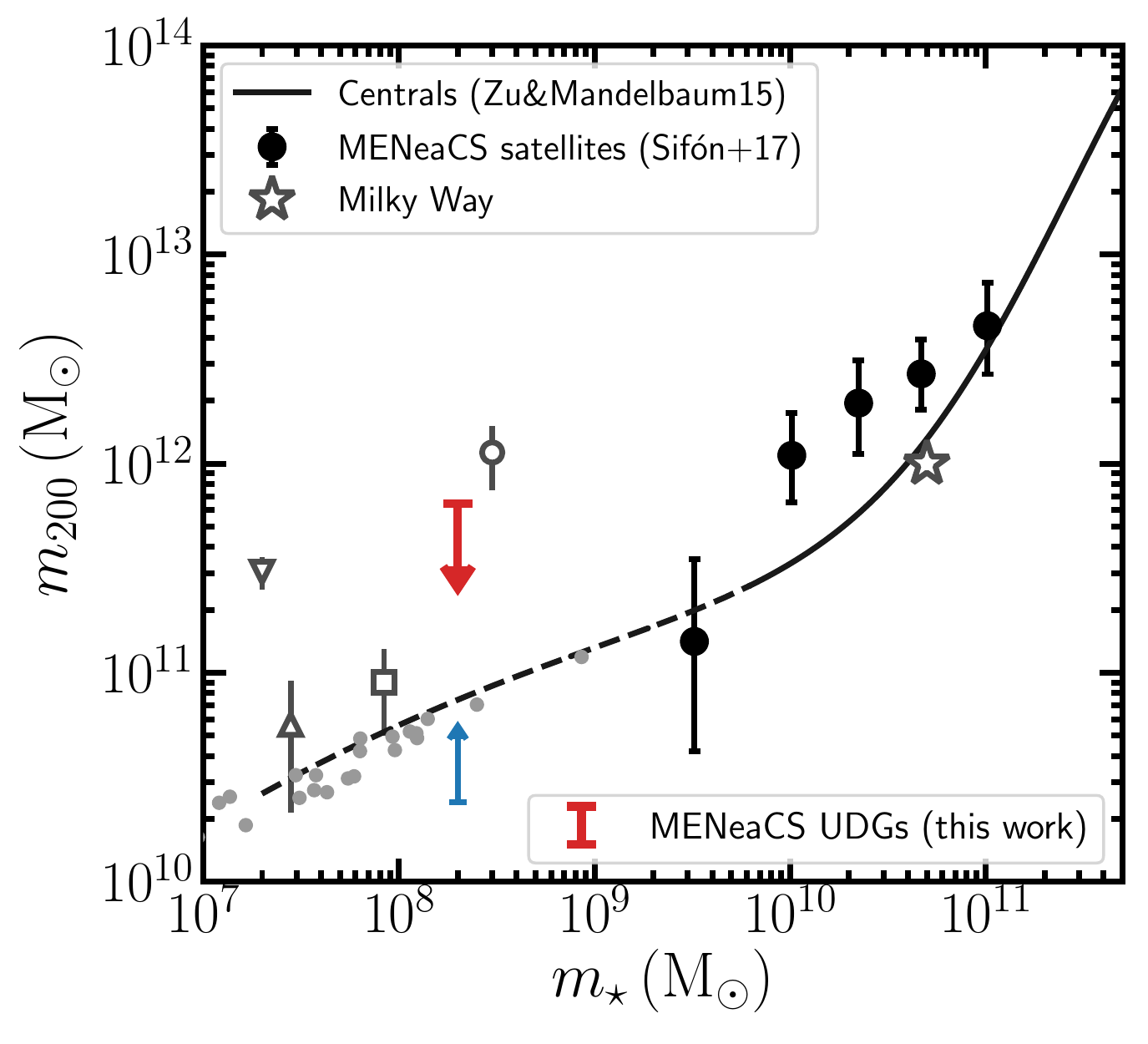}}\caption{Weak lensing constraints on the average halo mass of MENeaCS UDGs, expressed as $m_{200}$, at 95\percent\ credibility. Symbols are as in \Cref{f:shmr30kpc}. We also show the total-to-stellar mass relation for central galaxies by \citet{zu15} and the approximate location of the Milky Way in the halo-stellar mass plane. Uncertainties for UDG masses from the literature have been scaled directly from those reported in each work and do not account for uncertainties associated with the extrapolation itself.}
\label{f:shmr200}
\end{figure}

To put our results in context with recent measurements of UDG masses, we also show in \Cref{f:shmr30kpc} extrapolations to 30 kpc of mass measurements of individual UDGs in the literature.\footnote{In all our extrapolations of published UDG masses, we assume an NFW density profile with a concentration $c=10$, consistent with the $c(M)$ relation of \cite{moline17} adopted in our analysis.} The three UDGs with published enclosed masses are DF 44, measured within 4.6 kpc using the stellar velocity dispersion \citep{vandokkum16}; VCC 1287\footnote{All stellar masses have been determined assuming a \citet{chabrier03} IMF except for that of VCC 1287, which \citet{beasley16_virgo} determined using a \citet{kroupa02} function. These two assumptions yield stellar masses typically within $<10$\percent\ of each other, and therefore do not impact our discussion.}, using the globular cluster velocity dispersion within 8.1 kpc \citep{beasley16_virgo}; and UGC 2162 within 5 kpc, using the width of the H \textsc{i} emission line \citep{trujillo17}. In addition, \cite{beasley16_acs} and \cite{peng16} estimated a virial mass $m_{200}\sim(9\pm4)\times10^{10}$ for the Coma UDG DF17 using an empirical scaling relation based on the number of globular clusters. With the exception of DF44, all UDGs have measured masses smaller than our upper limit, as expected. \Cref{f:shmr30kpc} shows that DF44 is a mild outlier in terms of halo mass compared to its stellar mass. While the stellar mass of DF44, $\mstar\sim3\times10^8\,\Msun$, is similar to the average stellar mass in our sample, it has an effective radius $\reff=4.5\,\mathrm{kpc}$, significantly larger than the average $\reff$ of our sample. In fact, DF44 is among the largest UDGs from a sample of more than 800 discovered in Coma \citep{koda15}, so it is not entirely surprising that it may be an outlier in the halo mass-stellar mass plane, and suggests that---as with any galaxy sample---there is a large scatter in the halo masses of UDGs. We also show the masses of UDGs in the NIHAO simulations \citep{wang15}, as shown in \cite{dicintio17}, which seem to follow the total-to-stellar mass relation of regular satellites by \cite{sifon17_meneacs} remarkably well. These simulated UDGs exist in isolation, not in clusters, and the negligible scatter in the total-to-stellar mass relation shown by NIHAO UDGs---which are the result of gas-driven outflows spreading the stars of the galaxy---suggests that the UDG population is more diverse than captured by the simulations.

\cite{vdburg16} have shown that UDGs follow the same radial distribution within clusters as regular and compact dwarf galaxies down to $r\sim0.15r_{200}\approx300\,\mathrm{kpc}$. The fact that they are able to survive so close to the centres of massive clusters suggests that they must have total masses greatly exceeding their visible stellar mass. They derived a rough lower limit of the average UDG subhalo mass of $m(<6\,\mathrm{kpc})\gtrsim2\times10^9\,\Msun$. We show this lower limit extrapolated to 30 kpc as a blue arrow in \Cref{f:shmr30kpc}; it is entirely consistent with the upper limit implied by our lensing measurements. Together, these two limits bracket the average UDG mass to a range of approximately an order of magnitude. We caution that this ``tidal argument'' lower limit is not as precise as our weak lensing upper limit since it makes several simplifying assumptions (e.g., it assumes static satellites), but it serves as an important consistency check.

\section{Discussion}
\label{s:discussion}

Compared to stellar dynamics, lensing has the advantage that it probes the total masses of galaxies, as opposed to the masses within the small radii within which velocity dispersions can be measured. Therefore, lensing allows for a more straightforward comparison with theoretical predictions, without the need for an extrapolation of the density profile from $\lesssim10$ kpc to $\gtrsim100$ kpc (although lensing also relies to some extent on an assumed density profile). Weak lensing has however the disadvantage that the total and stellar masses are measured at different radii, and the question of how dark matter--dominated these galaxies are, compared to dwarf or `normal' galaxies, is difficult to address.

Our weak lensing measurements suggest the typical UDG (with $\langle\mstar\rangle=2\times10^8\,\Msun$, $\langle\reff\rangle=2.8$ kpc and $n_\mathrm{Sersic}=1.7$) in a massive galaxy cluster resides in a dark matter halo whose effective virial mass is at most half the mass of the Milky Way. We show this comparison explicitly in \Cref{f:shmr200}, where we further extrapolate the UDG masses plotted in \Cref{f:shmr30kpc} to their corresponding $r_{200}$ to obtain effective $m_{200}$ values, and show lensing measurements of $m_{200}$ for \meneacs\ satellites from \cite{sifon17_meneacs} and UDGs in our sample (we remind the reader that we use the word `effective' to highlight the fact that $m_{200}$ is physically ill-defined for subhaloes, but it provides a convenient mathematical parameter). These effective virial masses, or $m_{200}$, can be viewed as a rough estimate of the halo mass of the galaxy at the time of infall, assuming that tidal stripping has affected primarily the outer regions of galaxies. We also show for context the total-to-stellar mass relation measured for central galaxies by \cite{zu15}. Although there is a large extrapolation to low masses to allow a comparison with UDGs, the extrapolated curve is consistent with other measurements that extend closer to $m_\star\simeq10^9\,\Msun$ \citep[e.g.,][]{leauthaud12}, and potential errors of a factor of a few in the relation are not relevant for this first comparison. At low masses uncertainties are still large, but the total-to-stellar mass relations of centrals and satellites are roughly within a factor two of each other \citep{sifon17_meneacs}, and so is the relation for UDGs. The large statistical uncertainties preclude more detailed comparisons at the time.

Our measurement of the \emph{average} halo mass of UDGs helps bring the previous individual measurements within a broader context. Our results are broadly consistent with scenarios of UDG formation via internal processes, where UDGs occupy the extreme end of a continuous distribution of dwarf galaxies.  We cannot however constrain the extent of this distribution (i.e., its scatter) in terms of total mass. It is furthermore plausible that the scatter in the total-to-stellar mass relation of UDGs correlates with other physical properties, most notably the extent of the stellar light \citep[i.e., $\reff$, as suggested by][]{zaritsky17}, and it may well be the case that extreme examples such as DF44 be failed $L^*$ galaxies as suggested by \cite{vandokkum15_spec}. However, it would be difficult to extract information about the formation of UDGs from these extreme objects. Because UDGs can only be reliably identified in low-redshift clusters, the sample of UDGs is not likely to increase significantly. Instead, more sensitive UDG lensing measurements will require a significantly higher source density, which can only be obtained with the \textit{Hubble} Space Telescope. Such measurements will allow us to place more stringent constraints on the average masses of UDGs and move toward a physical understanding of their origin.

\section*{Acknowledgements}

We thank Pieter van Dokkum for valuable discussions and suggestions, Arianna Di Cintio for sharing the masses of NIHAO UDGs and Nicola Amorisco, Javier Rom\'an, and Ying Zu for useful comments on the manuscript. CS, HH and RH acknowledge support from the European Research Council under FP7 grant number 279396. RFJvdB acknowledges support from the European Research Council under FP7 grant number 340519.

Based on observations obtained with MegaPrime/MegaCam, a joint project of CFHT and CEA/DAPNIA, at the Canada-France-Hawaii Telescope (CFHT) which is operated by the National Research Council (NRC) of Canada, the Institute National des Sciences de l'Univers of the Centre National de la Recherche Scientifique of France, and the University of Hawaii.

\bibliographystyle{mnras}
\bibliography{bibliography}

\end{document}